\documentstyle[12pt]{article}
\begin{document}
\catcode`@=11
\newif\if@defeqnsw \@defeqnswtrue

\def\eqnarray{\stepcounter{equation}\let\@currentlabel=\theequation
\if@defeqnsw\global\@eqnswtrue\else\global\@eqnswfalse\fi
\global\@eqnswtrue
\tabskip\@centering\let\\=\@eqncr
$$\halign to \displaywidth\bgroup\hfil\global\@eqcnt\z@
  $\displaystyle\tabskip\z@{##}$&\global\@eqcnt\@ne
  \hfil$\displaystyle{{}##{}}$\hfil
  &\global\@eqcnt\tw@ $\displaystyle{##}$\hfil
  \tabskip\@centering&\llap{##}\tabskip\z@\cr}

\def\yesnumber{\global\@eqnswtrue}

\def\@@eqncr{\let\@tempa\relax\global\advance\@eqcnt by \@ne
    \ifcase\@eqcnt \def\@tempa{& & & &}\or \def\@tempa{& & &}\or
     \def\@tempa{& &}\or \def\@tempa{&}\else\fi
     \@tempa \if@eqnsw\@eqnnum\stepcounter{equation}\fi
     \if@defeqnsw\global\@eqnswtrue\else\global\@eqnswfalse\fi
     \global\@eqcnt\z@\cr}


\def\@eqnacr{{\ifnum0=`}\fi\@ifstar{\@yeqnacr}{\@yeqnacr}}

\def\@yeqnacr{\@ifnextchar [{\@xeqnacr}{\@xeqnacr[\z@]}}

\def\@xeqnacr[#1]{\ifnum0=`{\fi}\cr \noalign{\vskip\jot\vskip #1\relax}}

\def\eqalign{\null\,\vcenter\bgroup\openup1\jot \m@th \let\\=\@eqnacr
\ialign\bgroup\strut
\hfil$\displaystyle{##}$&$\displaystyle{{}##}$\hfil\crcr}
\def\endeqalign{\crcr\egroup\egroup\,}


\def\cases{\left\{\,\vcenter\bgroup\normalbaselines\m@th \let\\=\@eqnacr
    \ialign\bgroup$##\hfil$&\quad##\hfil\crcr}
\def\endcases{\crcr\egroup\egroup\right.}


\def\eqalignno{\stepcounter{equation}\let\@currentlabel=\theequation
\if@defeqnsw\global\@eqnswtrue\else\global\@eqnswfalse\fi
\let\\=\@eqncr
$$\displ@y \tabskip\@centering \halign to \displaywidth\bgroup
  \global\@eqcnt\@ne\hfil
  $\@lign\displaystyle{##}$\tabskip\z@skip&\global\@eqcnt\tw@
  $\@lign\displaystyle{{}##}$\hfil\tabskip\@centering&
  \llap{\@lign##}\tabskip\z@skip\crcr}

\def\endeqalignno{\@@eqncr\egroup
      \global\advance\c@equation\m@ne$$\global\@ignoretrue}


\@namedef{eqalignno*}{\@defeqnswfalse\eqalignno}
\@namedef{endeqalignno*}{\endeqalignno}


\def\eqaligntwo{\stepcounter{equation}\let\@currentlabel=\theequation
\if@defeqnsw\global\@eqnswtrue\else\global\@eqnswfalse\fi
\let\\=\@eqncr
$$\displ@y \tabskip\@centering \halign to \displaywidth\bgroup
  \global\@eqcnt\m@ne\hfil
  $\@lign\displaystyle{##}$\tabskip\z@skip&\global\@eqcnt\z@
  $\@lign\displaystyle{{}##}$\hfil\qquad&\global\@eqcnt\@ne
  \hfil$\@lign\displaystyle{##}$&\global\@eqcnt\tw@
  $\@lign\displaystyle{{}##}$\hfil\tabskip\@centering&
  \llap{\@lign##}\tabskip\z@skip\crcr}

\def\endeqaligntwo{\@@eqncr\egroup
      \global\advance\c@equation\m@ne$$\global\@ignoretrue}

\@namedef{eqaligntwo*}{\@defeqnswfalse\eqaligntwo}
\@namedef{endeqaligntwo*}{\endeqaligntwo}

%
%
%
%
%
\newtoks\@stequation

\def\subequations{\refstepcounter{equation}%
  \edef\@savedequation{\the\c@equation}%
  \@stequation=\expandafter{\theequation}
  \edef\@savedtheequation{\the\@stequation}
  \edef\oldtheequation{\theequation}%
  \setcounter{equation}{0}%
  \def\theequation{\oldtheequation\alph{equation}}}

\def\endsubequations{%
  \setcounter{equation}{\@savedequation}%
  \@stequation=\expandafter{\@savedtheequation}%
  \edef\theequation{\the\@stequation}%
  \global\@ignoretrue}


\def\big#1{{\hbox{$\left#1\vcenter to1.428\ht\strutbox{}\right.\n@space$}}}
\def\Big#1{{\hbox{$\left#1\vcenter to2.142\ht\strutbox{}\right.\n@space$}}}
\def\bigg#1{{\hbox{$\left#1\vcenter to2.857\ht\strutbox{}\right.\n@space$}}}
\def\Bigg#1{{\hbox{$\left#1\vcenter to3.571\ht\strutbox{}\right.\n@space$}}}

\catcode`@=12
%
%
\newcommand{\gsimeq}{\mathop{>}\limits_{\displaystyle{\sim}}}
\newcommand{\lsimeq}{\mathop{<}\limits_{\displaystyle{\sim}}}
\newcommand{\sigmin}{\sigma_{\gamma \gamma}^{\rm min}}
\newcommand{\as}{\alpha_s}
\newcommand{\amu}{A_t + \mu \cot \! \beta}
\newcommand{\mgg}{M_{\gamma \gamma}}
\newcommand{\fourl}{l^+ l^+ l^- l^-}
\newcommand{\fourtau} {\tau^+ \tau^+ \tau^- \tau^-}
\newcommand{\tht}{\theta_t}
\newcommand{\tanb}{\tan \! \beta}
\newcommand{\msb}{m_{\tilde{b}_L}}
\newcommand{\msbsq}{m^2_{\tilde{b}_L}}
\newcommand{\mt}{m_t}
\newcommand{\mtsq}{m^2_t$}
\newcommand{\mst}{m_{\tilde{\tau}}}
\newcommand{\msti}{m_{\tilde{\tau}_1}}
\newcommand{\mstii}{m_{\tilde{\tau}_2}}
\newcommand{\mstl}{m_{\tilde{\tau}_L}}
\newcommand{\mstr} {m_{\tilde{\tau}_R}}
\newcommand{\msl} {m_{\tilde{l}}}
\newcommand{\mstlsq}{m^2_{\tilde{\tau}_L}}
\newcommand{\mstrsq}{ m^2_{\tilde{\tau}_R}}
\newcommand{\mstlr}{m_{\tilde{\tau}_{L,R}}}
\newcommand{\mstlrsq}{ m^2_{\tilde{\tau}_{L,R}}}
\newcommand{\mstssq}{ m^2_{\tilde{\tau}_1}}
\newcommand{\msttsq}{ m^2_{\tilde{\tau}_2}}
\newcommand{\stst}{\tilde{t}_1 \tilde{t}_1^*}
\newcommand{\sigst} { \sigma_{\tilde{t}_1} }
\newcommand{\msig} { m_{\sigma_{\tilde t}} }
\newcommand{\gamgam} { \gamma \gamma }
\newcommand{\selr} { \tilde{e}_{L,R} }
\newcommand{\smulr} { \tilde{\mu}_{L,R} }
\newcommand{\se} { \tilde{e}}
\newcommand{\smu} { \tilde{\mu}}
\newcommand{\sti} { \tilde{\tau}_1 }
\newcommand{\stii} { \tilde{\tau}_2 }
\newcommand{\thetat}{\theta_{\tau}}
\newcommand{\cost}{\cos\theta_{\tau}}
\newcommand{\costs}{\cos^2\theta_{\tau}}
\newcommand{\sint}{\sin\theta_{\tau}}
\newcommand{\sints}{\sin^2\theta_{\tau}}
\newcommand{\sw}{\sin\!\theta_W}
\newcommand{\cw}{\cos\!\theta_W}
\newcommand{\tw}{\tan\!\theta_W}
\newcommand{\brchi}{Br(\sti\rightarrow \chi_1^0 \tau)}
\newcommand{\gami}{\Gamma(\sti\rightarrow\chi_1^0\tau)}
\newcommand{\gamii}{\Gamma(\sti\rightarrow\chi_1^0\tau)}
\newcommand{\gamci}{\Gamma(\sti\rightarrow\chi_1^-\tau)}
\newcommand{\cosb}{\cos\!\beta}
\newcommand{\sinb}{\sin\!\beta}
\newcommand{\sws}{\sin^2\theta_W}
\newcommand{\cws}{\cos^2\theta_W}
\newcommand{\tws}{\tan^2\theta_W}
\newcommand{\mtau}{m_{\tau}}
\newcommand{\chio}{ \chi^0_1}
\newcommand{\chii}{\chi^0_i}
\newcommand{\chimi}{\chi^-_i}
\newcommand{\aiiR}{a_{1i}^R}
\newcommand{\aiiL}{a_{1i}^L}
\newcommand{\st}{\tilde{\tau}}
\newcommand{\stl}{ \tilde{\tau}_L}
\newcommand{\str}{ \tilde{\tau}_R}
\newcommand{\mchi}{m_{{\chi}^0_1}}
\newcommand{\mchisi}{m_{{\chi}^0_i}}
\newcommand{\mchii}{m_{{\chi}^0_2}}
\newcommand{\ptau}{ P_{\tau}}
\newcommand{\ww}{ W^+W^- }
\newcommand{\epem}{ e^+e^-}
\newcommand{\mat}{{\cal M}^2_{\tilde{t}}}
\newcommand{\etauc}{E_{\tau}^c}
\newcommand{\ptauc}{p_{\tau}^c}
\newcommand{\vstau}{v_{\sti}}
\newcommand{\etaucs}{E_{\tau}^{c 2}}
\newcommand{\etaumax}{E_{\tau}^{\rm max}}
\newcommand{\etaumin}{E_{\tau}^{\rm min}}
\newcommand{\be}{\begin{equation}}
\newcommand{\ee}{\end{equation}}
\newcommand{\een}{\end{subequations}}
\newcommand{\ben}{\begin{subequations}}
\newcommand{\beq}{\begin{eqalignno}}
\newcommand{\eeq}{\end{eqalignno}}
\renewcommand{\thefootnote}{\fnsymbol{footnote} }
\noindent
\begin{flushright}
KEK--TH--425\\
KEK Preprint 94--171\\
December 1994
\end{flushright}
\vspace{1.5cm}
\pagestyle{empty}
\begin{center}
{\Large \bf Polarization of $\tau$ lepton from
scalar tau decay \\ as a probe of neutralino mixing
}\\
\vspace*{5mm}
Mihoko M. Nojiri\footnote{E--mail: NOJIRIN@JPNKEKVX}\\
{\em Theory Group, KEK, Oho 1--1, Tsukuba, Ibaraki 305, Japan}
\end{center}

\begin{abstract}
The $\tau$ lepton arising from the scalar tau ($\st$) decay
is naturally polarized.  $\ptau$ depends on the left--right mixing of
the $\st$  and the gaugino--higgsino mixing of the neutralino.
The polarization $\ptau$ could be measured from the energy distribution of
the decay products of $\tau$ at future $\epem$ colliders.
A measurement of $\ptau$ and of the $\st$ production cross section  allows to
determine both these mixing angles.
\end{abstract}
\clearpage
\setcounter{footnote}{0}
\pagestyle{plain}
\setcounter{page}{1}
\section*{1) Introduction}

The Minimal Supersymmetric Standard Model (MSSM)\cite{1} is one of the most
promising candidates of the models beyond the standard model (SM)\cite{2}. It
predicts the existence of superpartners of SM particles below a few TeV to
remove quadratic divergences which appear in radiative corrections of the SM;
thus the model is free from the so--called hierarchy problem\cite{3} of GUT
models. It should be noted that the gauge couplings unify very precisely at
high energy scale in MSSM \cite{4}, consistent with SUSY GUT predictions.

If supersymmetry is unbroken, a particle and its superpartner have equal mass.
Since no superpartner has been found so far the symmetry must be broken rather
badly. Several mechanisms to break supersymmetry have been proposed in the
framework of supergravity and superstring models \cite{5}. These models
predict relations between various soft SUSY breaking parameters at the GUT
scale (or Planck scale); these are the soft scalar masses $m_i$, gaugino
masses $M_i$, and trilinear couplings $A_i$. The masses of SUSY particles at
the weak scale are obtained by evolving the GUT scale parameters using the
renormalization group equation (RGE) of the MSSM. Thus the observation
of superpartners at $\epem$ or hadron colliders should give
important information on the models that describe the origin of supersymmetry
breaking.

In this paper, we study the production and decay of the superpartner of the
tau ($\st$) at future $\epem$ colliders in the framework of the MSSM.
Production and decay of $\selr$ and $\smulr$ at such colliders have already
been discussed in various articles \cite{6}. $\selr$ ($\smulr$) is produced at
$\epem$ colliders by $s$ channel exchange of photon and $Z^0$ for $\smulr$,
and also by $t$ channel exchange of neutralinos for $\selr$. The produced
$\selr$ ($\smulr$) then  decays into the lightest neutralino $\chio$, which is
a mixture of gauginos and higgsinos, plus an $e$ ($\mu$) through the gaugino
component of the neutralino,  giving a signal with leptons and missing
momentum.

$\st$ production and decay is different from that of $\tilde{e}$ and
$\tilde{\mu}$ because the $\tau$ lepton has a non--negligible Yukawa coupling.
Due to this coupling, $\str$ and $\stl$ mix, thus the mass eigenstates are not
necessarily current eigenstates \cite{7}. The same Yukawa coupling appears as
a non--negligible $\tau \st \tilde{H_1^0}$ coupling, where $\tilde{H_1^0}$ is
a neutral higgsino\cite{8}. This interaction is involved in $\st$ decay
processes, since the $\chi$'s are mixtures of higgsinos and gauginos. Another
feature of $\st$ decay that distinguishes it from other slepton decays is that
the $\tau$ lepton arising from  the decay $\st\rightarrow\chi\tau$  decays
further in the detector, which enables us to measure the average polarization
of the $\tau$ ($\ptau$) \cite{9,10}. The purpose of this paper is to point out
that one can measure the $\stl-\str$ mixing of $\sti$ {\em and} the
higgsino--gaugino mixing of $\chi_i^0$ by measuring both the cross section for
$\st$ production and the average polarization of $\tau$ from $\st$ decays.

The rest of this paper is organized as follows. In section 2 we discuss
$\sti\sti^*$ production at an $\epem$ collider with $\sqrt{s}=500$ GeV. We
point out that the cross section for $\sti$ production depends on $\msti$ and
on the $\sti$ left--right mixing angle $\thetat$. If a polarized electron beam
is available, $\thetat$ can be determined from  the cross section, assuming
$\msti$ is known (e.g. from kinematical distributions).

In section 3 we show that the $\tau$ lepton arising from $\sti$ decay is
naturally polarized. The $P_{\tau}(\tilde{\tau_1}\rightarrow\chi_i^0\tau)$
depends on both $\thetat$ and the neutralino mixing. This latter dependence
arises because the interaction of gauginos with (s)fermions preserves
chirality, while the interaction of higgsinos flips it. Removing the $\thetat$
dependence by using the measured value of $\thetat$, one can isolate
information on the neutralino mixing. We also show that the branching ratio of
the decay mode $\sti\rightarrow\chio\tau$, which provides a good signal for
$\sti$ production, is sizable in a wide region of the parameter space.

$\ptau$ can be measured through the energy distribution of the decay products
of the polarized tau lepton. A complication arises since this distribution
also depends on the energy distribution of the parent tau coming from $\sti$
decay. In section 4 we describe the dependence of the energy distribution on
$\mst$, $\mchi$ and $\ptau$. Possible backgrounds are also discussed here.

Section 5 is devoted to a discussion and summary. We also describe  the
predictions for the mass spectrum in the minimal supergravity model, and its
implications.




\section*{2) The cross section for $\sti\sti^*$ production}
The scalar tau lepton $\tilde{\tau}_{L(R)}$ is the superpartner of
$\tau_{L(R)}$. $\str$ is an SU(2) singlet and has hypercharge $-1$, while
$(\tilde{\nu}_{\tau},$$\tilde{\tau})_L$ is a SU(2) doublet and has
hypercharge $-1/2$. The mass matrix of $\st$ can be written as follows
\cite{7,16},
\be\label{e1}
{\cal M}^2_{\st}=\left(\begin{array}{cc}m_{LL}^2 & m_{LR}^2\\
m_{LR}^2& m_{RR}^2\end{array}\right)
=\left( \begin{array}{cc}
m_L^2 + m_{\tau}^2 + 0.27 D &   -m_{\tau}(A_{\tau} + \mu \tan\beta)\\
 -m_{\tau}(A_\tau + \mu \tan\beta)&  m_R^2 +m_{\tau}^2 + 0.23D\\
 \end{array}\right)
\begin{array}{c}\stl\\\str\end{array}
\ee
where $m_R $ and $m_L$ are soft breaking scalar mass parameters of $\str$ and
$(\tilde{\nu}_{\tau},\st)_L$, $A_{\tau}$ is a trilinear coupling of
$\st_L\st_R H_1$, and $\mu$ is the Higgsino mass parameter. $\tan\beta=
\langle H^0_1\rangle/ \langle H^0_2\rangle$ is the ratio of the vacuum
expectation values of the two Higgs doublets of the MSSM, and  $D=-m_Z^2
\cos(2\beta)$. $\str$ and $\stl$ then mix to form two mass eigenstates $\sti$
and $\stii$ ($\msti<\mstii$):
\be\label{e2}
\left(\begin{array}{c} \sti\\\stii\end{array}\right)
=\left(\begin{array}{cc}\cost &\sint\\
 -\sint&\cost\end{array}\right)
\left(\begin{array}{c} \stl\\\str\end{array}\right).
\ee

At future $\epem$ linear colliders such as JLC, it is proposed to have a
highly polarized electron beam\cite{12}. At the Stanford Linear
Collider (SLC), an average polarization of 80\% has already been achieved in
the 1994 run. At future linear colliders, a beam polarization of more than
90\% is expected\cite{14,11}. Having a polarized beam also helps to reduce
background from $W^+W^-$ pair production, which makes SUSY analyses
considerably easier\cite{11}. Thus we give the differential and total
cross sections ($d\sigma/d\cos\theta_{\tau},\sigma$) for $\sti$ production by
$\epem$ annihilation after taking spin sum and average only for the positron.
$\st_1\st^*_1$ production occurs through s-channel exchange of $\gamma$ and
$Z^0$:
\ben\label{e3}\beq
\frac{d\sigma
}{d\cos\theta}(\epem\rightarrow \sti\sti^*)
&=\frac{\beta_f}{32\pi s}\frac{1}{2}
\vert T(\epem\rightarrow \sti\sti^*)
\vert^2 \sin^2\theta;
\label{e3a}
\\
\sigma
(\epem\rightarrow \sti\sti^*)
&=
\frac{\beta_f}{3\cdot 16\pi s}\frac{1}{2}\vert T(\epem\rightarrow \sti\sti^*)
\vert^2;
\label{e3b}
\eeq\een
where
\beq
T(\epem\rightarrow \sti\sti^*) &= \sqrt{2} s \beta_f
\left[-g_Z^2 [\costs (\frac{1}{2}+\sws)+\sws\sints]
\right.\nonumber\\
\ \ \ \ \ \ \ &\cdot\left.
\frac{1}{s-m_Z^2 +i\Gamma_Z m_Z}\times
\left(\begin{array}{c}
\sws \\
-\frac{1}{2}+\sws
\end{array}
\right)
\begin{array}{c}
h_e=1/2\\
h_e=-1/2
\end{array}
\right. \nonumber\\
&\ \ \ \ \ \ \left. -e^2\frac{1}{s}\right].\label{e3p}
\eeq
Here $h_e$ is the helicity of the initial electron, and
$\beta_f=\sqrt{1-4\msti^2/s}$.\footnote{For the expression for
$\stii\stii^*$ production, replace $\cost$ and $\sint$ by $-\sint$ and
$\cost$.}

In the limit $s\gg m_Z$, Eq.(\ref{e3}) reduces to
\ben\label{e4}
\beq
T(h_e=1/2)&=- \sqrt{2}\beta_f
g^2 \tan^2\theta_W\left(\frac{1}{2}\costs+\sints\right);
\label{e4a}\\
T(h_e =-1/2)&=-\frac{ \sqrt{2}\beta_f g^2}{4}
 \left[ \left(1+\tan^2\theta_W\right)\costs
+2\tan^2\theta_W\sints\right].
\label{e4b}
\eeq
\een
If $h_e=1/2$ (or $P_e=1$), only $U(1)$ gauge boson exchange survives, thus the
 cross section for $\str\str^*$  production ($\cost=0$) is 4 times larger than
that for $\stl\stl^*$ ($\cost=1$) as one can see from Eq.(\ref{e4a}) and
(\ref{e3b})\cite{11}. On the other hand, due to the contribution of the
exchange of the $SU(2)$ gauge boson, the cross section for $\stl\stl^*$
production is larger than that for $\str\str^*$ if $h_e=-1/2$.

In fig.1, we show the $\thetat$ dependence of the  cross section for
$\sti\sti^*$ production. We take $\sqrt{s}=500$ GeV and $\msti=150$ GeV. The
solid curve is the cross section for $P_e=1$, and the dashed curve is for
$P_e=0$. If the electron beam is unpolarized, the cross section is about 0.06
pb and has very mild dependence on $\thetat$. Thus it is very difficult to
distinguish the value of $\thetat$ from the unpolarized cross section only.

On the other hand, for a right--handed electron beam $\sigma_{\str}\simeq
4\sigma_{\stl}$ as expected. In this case the minimum cross section for
$\sti\sti^*$ production is about 0.02 pb for the given choice of parameters.
${\cal L}=30 \ {\rm fb}^{-1}$  might be available per year at future $\epem$
colliders \cite{12}, which means at least six hundred $\st$ events can be used
for the analysis of the nature of $\sti$. If one knows $\msti$ very precisely,
e.g. from a threshold scan, one may extract $\thetat$ from the
measurement of the production cross section. Assuming all $\st$ production
events are identified, the cross section can be measured with statistical
error of less than 4\% for $\msti=150$ GeV. This corresponds to less than $\pm
9^{\circ}$ of uncertainty for $\thetat$. Notice that the soft breaking mass
parameters $m^2_{LL},\ m^2_{RR}, \ m^2_{LR}$ which appeared in Eq.(\ref{e1})
could be calculated if one further measures $m_{\tilde{\tau}_2}$; these
measurements give us a unique opportunity to extract the trilinear coupling
$A_{\tau}$ in $m_{LR}^2$.

How can $\sti$ production be identified? The lighter $\sti$ decays into
$\tau$ and a neutralino ($\chi^0_i$) or $\nu_{\tau}$ and a chargino
($\chi^-_i$). Here, $\chi^0_i$ is a mixture of $\tilde{B}$, $\tilde{W_3}$,
$\tilde{H_1}$ and $\tilde{H_2}$, while $\chi^-_i$ is a mixture of $\tilde W$
and $\tilde{H}$. The mass matrices of the neutralinos ($\chi^0_i$) and
charginos ($\chi^+_i$) are given as follows\cite{1}:
\newenvironment{arrayc}{\renewcommand{\arraycolsep}{0.cm}\begin{array}}
{\end{array}}
\ben\label{e9}
\beq
{\cal M}_N&(\tilde{B},\tilde{W_3},\tilde{H_1},\tilde{H_2})=
\nonumber\\
&\left(\begin{array}{cccc}
M_1 &0&-m_Z\sw\cosb&m_Z\sw\sinb\\
0 &M_2& m_Z\cw\cosb&-m_Z\cw\sinb\\
-m_Z\sw\cosb&m_Z\cw\cosb&0&-\mu \\
m_Z\sw\sinb&-m_Z\cw\sinb&-\mu & 0
\end{array}\right),\nonumber\\
\label{e9a}
\\
{\cal M}_c&(\tilde{W},\tilde{H})=\left(\begin{array}{cc}
M_2 & m_W \sqrt{2}\sinb\\
m_W\sqrt{2}\cosb & \mu
\end{array}
\right).
\label{e9b}
\eeq
\een
These mass matrices are diagonalized by a real orthogonal matrix $N$ for ${\cal
M}_N$, and unitary matrices $U$ and $V$ for ${\cal M}_c$ as follows:
\be
U^*{\cal M}_C V^{-1}=M_D^c, \ \ N {\cal M}_N N^{-1}=M_D^N.
\label{e10}
\ee
Unlike the notation of Haber and Kane\cite{1}, we assume $N$ to be real so
that $m_{\chi_i^0}$ can be either positive or negative; the sign must be kept
to understand later expressions. We take $\vert m_{\chi_1^0} \vert \leq
$$\vert m_{\chi_2^0} \vert \leq $$\vert m_{\chi_3^0} \vert \leq $ $\vert
m_{\chi_4^0} \vert$ and $0\leq m_{\chi_1^-} \leq $$m_{\chi_2^-}$. We assume
the mass relation $M_1=(5/3)\cdot \tws M_2$ for numerical calculations in
order to reduce the number of parameters.

In the MSSM, there is a multiplicatively conserved R parity, which is assigned
 $-1$ for superparticles (superpartners of fermions, Higgs bosons and gauge
bosons), and $+1$ otherwise. Therefore the decay of a superparticle creates at
least one superparticle, so that the lightest of them is stable. Because no
such particle has been found in cosmic rays, the lightest supersymmetric
particle (LSP) must be neutral and weakly interacting. The lightest neutralino
$\chio$  is generally considered to be the best candidate for the LSP; it
escapes detection in collider experiments because it is neutral and does not
decay inside detectors.

If both of the $\sti^{(*)}$ decay into $\chi^0_1$ and $\tau(\bar\tau)$, the
event should look like $\tau\bar{\tau}$ + missing momentum, as the two
$\chi^0_1$ escape from detection. This is the simplest event signature for
$\sti\sti^*$ production. If $\sti$ is the second lightest SUSY particle, all
$\sti$ decay through this mode. The tau leptons can be identified through
their decays into leptons or mesons, which we discuss in section 4.

If $\chi_i^0\tau$ or $\chi_i^-\nu$ modes are also kinematically accessible,
the signatures for $\sti$ production are more complicated. The branching ratios
into heavier neutralinos and charginos depend on $\thetat$, $M$, $\tan\beta$,
and $\mu$ very strongly, and could be dominant, as will be discussed in
section 3. If we assume $\sti$ is the lightest of all sfermions (which  means
$\chi$'s cannot decay into sfermions), those $\chi_i^0$ and  $\chi_i^-$ would
further decay as
\beq
\chii\rightarrow\chi^0_{i'} f\bar{f},\chi^{\pm}_{i'} f\bar{f'},
\nonumber\\
\chi^{\pm}_i\rightarrow\chi^0_{i'} f\bar{f'},\chi^{\pm}_{i'} f\bar{f}.
\eeq
When the decay cascade is terminated, the final state contains one $\chio$ due
to the R parity conservation, again giving missing momentum to the event. When
 both $\sti$ and $\sti^*$ decay into a neutralino and a tau, we get an event
$\tau\bar{\tau}$ + (jets and/or leptons) + missing momentum. On the other
hand, if both of $\sti$ decay into a chargino and $\nu_{\tau}$, the event does
not contain any $\tau$ lepton in the final state (unless they come from --ino
decays), so it is not a good $\sti$ signature.

One can in principle measure both $\msti$ and $\thetat$ by comparing the cross
section for $\sti$ production at $P_e=1$ and $P_e=0$, as $\sigma(P_e=0)$ is
quite insensitive to $\thetat$. However, the events at $P_e=0$ suffer a large
background from $W^+W^-$ production where both $W^+$ and $W^-$ decay into
$\tau\nu$. The production cross section of $W^+W^-$ is about 7 pb for $P_e=0$,
while it is only 0.05 pb for $P_e=1$ as a right handed electron does not
couple to transverse $W$ bosons \cite{11, 11b}. Notice that most $W$ bosons
will be emitted in the forward direction when $P_e=0$, while the signal tau
leptons tend to be emitted at large angles as one can see from Eq.(\ref{e3a}).
The  cross section for $W^+W^-$ production  at $P_e=0$ is reduced to 1 pb if
we require $\vert\cos\theta\vert<0.8$ for the azimuthal angle of the $W$
bosons \cite{12}. Then the background to the process $\st_1\st^*_1 \rightarrow
\tau\tau$ + missing momentum with a similar angle cut on the tau is roughly
$10$ fb and $0.5$fb for $P_e=0$ and $P_e=1$, respectively. Although an
accoplanarity angle cut reduces the background from $W^+W^-$ even further, the
measurement of $\sigma_{\sti}(P_e=0)$ is clearly more difficult than that of
$\sigma_{\sti}(P_e=1)$.

\section*{3) $\tau$ polarization in $\sti$ decay}
\subsection*{a) $\ptau(\sti\rightarrow\tau\chi^0_i)$}
The $\tau$ lepton arising from  the decay $\sti\rightarrow\tau\chi^0_i$
further decays into $A\nu_\tau$ where $A=e \nu, \pi,\rho,a_1...$. The decay
distributions of the $\tau$ decay products depend on the polarization of the
parent\cite{9}. In particular, for each decay channel the momentum
distribution of the decay products ($\pi^-$, $\rho\rightarrow \pi^-\pi^0$,...)
differs significantly according to whether the parent is $\tau_R^-$($h=1/2$)
or $\tau_L^- $($h=-1/2$). If the $\tau$ lepton is relativistic, $P_{\tau}$ can
then be determined from the energy distribution of the decay products
\cite{10}.
Compared to the case of $\tilde{e}$ and $\tilde{\mu}$ decays, where the
polarization of the final state lepton is not measurable, we therefore have
one more piece of information available for the study of the nature of $\sti$
and $\chi^0_i$. In this section, we discuss the decay width of $\tilde{\tau}$
into $\chi_i^0\tau$ and $\chi_i^-\nu_{\tau}$, while keeping our eyes on the
polarization of the $\tau$ lepton.

The stau-ino-fermion interaction relevant for $\sti$ decay is expressed by the
following Lagrangian\cite{8}:
\newenvironment{sarray}{\renewcommand{\arraystretch}{0.5}
\begin{array}}
{\end{array} \renewcommand{\arraystretch}{1}}
\be
{\cal L}=
\sum_{
\begin{array}{c}\scriptstyle i=1,2\\ \scriptstyle j=1,..,4 \end{array}}
\tilde{\tau}_i\  \bar{\tau}( P_L a^R_{ij}+P_R a^L_{ij})\chi^0_j
+ \sum_{  \begin{array}{c}\scriptstyle i=1,2\\ \scriptstyle j=1,2\end{array}}
\tilde{\tau}_i\  \bar{\nu}_{\tau} P_R b_{ij}\chi^+_j ,
\label{e5}
\ee
where
\ben\label{e6}\beq
&\left( \begin{array}{c}a^{R(L)}_{1j}\\ a^{R(L)}_{2j}\end{array}\right)
=
\left(\begin{array}{cc}\cost&\sint\\
-\sint&\cost \end{array}\right)
\left(\begin{array}{c}a^{R(L)}_{Lj}\\ a^{R(L)}_{Rj}\end{array}\right),
\nonumber
\\
&\left( \begin{array}{c}b_{1j}\\ b_{2j}\end{array}\right)
=
\left(\begin{array}{cc}\cost&\sint\\
-\sint&\cost \end{array}\right)
\left(\begin{array}{c}b_{Lj}\\ b_{Rj}\end{array}\right),
\\
a^R_{Lj}&=-\frac{g m_{\tau}}{\sqrt{2}m_W \cos\beta}N_{j3},
\ \
a^L_{Lj}=\frac{g}{\sqrt{2}}\left[ N_{j2}+N_{j1}\tan\theta_W\right],
\nonumber
\\
a^R_{Rj}&=-\frac{2g}{\sqrt{2}}N_{j1}\tan\theta_W,
\ \
a^L_{Rj}=-\frac{g m_{\tau}}{\sqrt{2}m_W \cos\beta}N_{j3},
\nonumber\\
b_{Lj}&=-gU_{j1},
\ \
b_{Rj}=\frac{gm_{\tau}}{\sqrt{2}m_W \cos\beta } U_{j2}.
\eeq
\een
The coefficients of the terms which flip chirality of the (s)fermion
($a^L_{Ri}, a^R_{Li}$) are proportional to the $\tau$ Yukawa coupling
$gm_{\tau}/(m_W \cos\beta\sqrt{2})$. Notice that this coupling is proportional
to $\cos^{-1}\beta$, which is close to $\tan\beta$ when $\tanb\gg 1$.

The decay width of the process $\sti\rightarrow \chi^0_i\tau $
is expressed as
\beq\label{e8}
\Gamma(\sti\rightarrow \chi^0_i\tau )=
\frac{\beta_f}{16\pi \mst}\vert{\cal M}(\sti\rightarrow \chi^0_i\tau )
\vert^2,
\eeq
where
\ben\label{e7}
\beq
{\cal M}\left(\sti\rightarrow \chi^0_i(\bar{h})\tau(h)\right)&=\bar{u}(\tau,h)
(P_La^R_{1i}+P_R a^L_{1i}) v(\chi^0,\bar{h})\nonumber\\
&=\frac{1}{2}\left[(a^R_{1i}+a^L_{1i})H_s +(a^L_{1i}-a^R_{1i})H_p\right];
\label{e7a}
\\
H_s&=\sqrt{\msti^2-(\mchisi+\mtau)^2}\ \delta_{\lambda_f 0},\label{e7b}\\
H_p&=\sqrt{\msti^2-(\mchisi-\mtau)^2}\
\delta_{\lambda_f 0}(-)^{\bar{h}+\frac{1}{2}}.\label{e7c}
\eeq
\een
Here $h$ and $\bar{h}$ are helicities of $\chi_i^0$ and $\tau$, respectively,
and $\lambda_f\equiv h-\bar{h}$. Recall that $\mchisi$ can have either sign
here. The decay width $\Gamma(\sti\rightarrow\nu_{\tau}\chi^-_i)$ can be
obtained by replacing $a^L_{1i}\rightarrow b_{1i}$ $a_{1i}^R\rightarrow 0$,
$m_{\chi_i^0} \rightarrow m_{\chi_i^-}$ and $m_{\tau}$ into zero. In this case
$H_s=H_p(-)^{\bar{h}+1/2}$, thus $\sti$ decays only to a $\bar{h}=-1/2$ (left
handed) neutrino state as it should be.

In the limit $m_{\tau}\ll\msti$, the difference between $|H_s|$ and $|H_p|$ is
small. In such a case $\Gamma(\sti\rightarrow \chii \tau_R) \propto
(a_{1i}^R)^2$ and $\Gamma(\chii\tau_L)\propto (a_{1i}^L)^2$. The average
polarization of the tau leptons is expressed in this limit as
\be\label{e11}
\ptau=\frac{Br(\chi_i^0\tau_R)-Br(\chi_i^0\tau_L)
}{Br(\chi_i^0\tau_R)+Br(\chi_i^0\tau_L)}=
\frac{(\aiiR )^2 -(\aiiL )^2}{(\aiiR )^2 +(\aiiL )^2}.
\ee
$\ptau$ depends on both the stau mixing $\thetat$ and on the neutralino mixing
$N_{ij}$. In the limit where $\chii$ is gaugino--like, only $N_{i1}$ and
$N_{i2}$ are sizable. This leads to $ \ptau=-1$ for $\cost=1$, while $
\ptau=+1$ for $\cost=0$ from Eq.(\ref{e11}) and (\ref{e6}). This results from
the supersymmetric version of the chirality conservation of gauge
interactions. On the other hand, if $\chii$ is higgsino--like, one finds
$\ptau=+1$ for $\cost=1$ and $\ptau=-1$ for $\cost=0$, because the
supersymmetric version of Yukawa interactions flips the chirality of a
(s)fermion. In section 2, we have discussed the measurement of $\thetat$ from
the cross section for $\st\st^*$ production; even if $\sti$ is a mixed state
of $\stl$ and $\str$, one can still extract the information about neutralino
mixing from  the measured $\ptau$ and $\thetat$ by solving Eq.(\ref{e11}).

To be more specific, the expression for
$\ptau$ for $\str$ decay is 
\be
\ptau(\str\rightarrow\chio\tau)
=\frac{(\sqrt{2}N_{11}\tw)^2-\left( \frac{m_{\tau}}
{\sqrt{2}m_W\cosb} N_{13}\right)^2}
{(\sqrt{2}N_{11}\tw)^2+\left( \frac{m_{\tau}}
{\sqrt{2}m_W\cosb} N_{13}\right)^2}.
\label{e15}\ee
In fig. 2 we show contours of fixed $\ptau (\str\rightarrow \chio \tau)$ in
the $M_1-\mu$ plane for $\tanb=10$ (fig. 2a) and $\tanb=2$ (fig. 2b). $\ptau$
decreases monotonically as $M_1$ increases for a fixed value of $\mu$, as
$\chio$ is bino--like ($N_{11}\simeq 1$) for $M_1\ll\vert\mu\vert$, while
$\chio$ is higgsino--like if $M_1\gg\vert\mu\vert$ ($N_{11}\ll 1$). However,
as the Yukawa coupling of $\tau$ is small ($\propto m_{\tau}/(m_W\cdot
\cosb$)), $\st_R$ still dominantly decays into $\tau_R$ after $M_1$ becomes
larger than $\vert\mu\vert$. $\tau$ becomes unpolarized only when $M_1\simeq
260$ GeV for $\mu=-100$ GeV and $\tanb=10$. For $\tanb=2$ and the same value
of $\mu$, $\ptau$ becomes zero when $M_1\simeq 500$ GeV due to the even
smaller Yukawa coupling.

Although it is true that a gaugino--(higgsino--)like neutralino tends to give
$\ptau\simeq +1 \ (-1)$, the actual value of $\ptau$ depends strongly on
$\tan\beta$. The polarization in the region where $\vert\mu\vert\ll M_1$ may
be understood from the perturbative diagonalization of the neutralino mass
matrix. Assuming that the differences between the diagonal elements is larger
than the off--diagonal elements, the two higgsino--like neutralino masses and
eigenvectors are given approximately by \cite{15}
\ben \label{e13} \beq
m_{\chi_{H}} &= \mu (1 + \delta); \nonumber \\
N_{Hi} &= \frac {1} {\sqrt{2}} \left( \frac { M_Z \sw
(\cosb+\sinb) } {M_1 - \mu}, \frac{ M_Z \cw (\cosb+\sinb)} {\mu - M_2}, 1, -1
\right) ; \label{e13a} \\
m_{\chi_{H'}} &= -\mu (1 + \delta'); \nonumber \\
N_{H'i} &= \frac {1} {\sqrt{2}} \left( \frac { M_Z \sw
(\cosb-\sinb) } {M_1 + \mu}, \frac{ M_Z \cw (\sinb-\cosb)} {M_2 + \mu}, 1, 1
\right) ; \label{e13b} \eeq \een
where
\ben \label{e14} \beq
\delta &= \frac {M_Z^2 ( 1 + \sin \! 2 \beta) } { 2 \mu} \left(
\frac {\sin^2 \! \theta_W} { \mu - M_1} + \frac {\cos^2 \! \theta_W} {\mu -
M_2}
\right) , \label{e14a} \\
\delta' &= \frac {M_Z^2 ( 1 - \sin \! 2 \beta) } { 2 \mu} \left(
\frac {\sin^2 \! \theta_W} { \mu + M_1} + \frac {\cos^2 \! \theta_W} {\mu +
M_2}
\right). \label{e14b}
\eeq \een
For $\vert M_1\vert \gg\vert\mu\vert$, one finds $\chi_1=\chi_{H'}$, $\chi_2=
\chi_{H}$ for $\mu<0$, while $\chi_1=\chi_H$ and $\chi_2=\chi_{H'}$ for
$\mu>0$.

Inserting Eq.(\ref{e13}) into Eq.(\ref{e15}), we estimate that $\ptau
(\st_R\rightarrow \chi_1\tau)=0$ at $M_1\simeq 280$ GeV for $\mu=-100$ GeV and
$\tanb=10$, which  roughly agrees with the numerical result shown in the
figure. For $\tanb=2$, the value of $M_1$ that leads to $\ptau=0$ is estimated
to be about 550 GeV.

Fig. 2b shows quite asymmetric behavior of the $\tau$ polarization for the
sign of $\mu$. This is due to the fact that the bino component of $\chio$ is
enhanced (suppressed) by the factor $\cosb\pm\sinb$ for $\mu>0 \ (\mu<0)$. For
large $\tanb$, this becomes irrelevant as $\cosb\ll\sinb$.

\subsection*{3b) The branching ratio into the lightest neutralino}
As discussed in section 2, the decay mode $\st_1\rightarrow \tau\chi^0_1$
gives the simplest signal for $\st_1$ production, while the modes
$\sti\rightarrow\nu\chi^-_i, \ \tau\chi^0_j(j>2)$ are more challenging. In
this subsection, we are going to see how big the branching ratio into the
lightest neutralino is among the other branching ratios.

In fig. 3, we show the $M_1$ dependence of the branching ratios of $\sti$, for
$\mu=-300$ GeV and $\msti=150$ GeV. In the region of parameter space shown in
this figure, $\chio\simeq\tilde{B}$, $\chi_2^0\simeq \tilde{W_3}$ and
$\chi_1^- \simeq\tilde{W}$, and only the decays into those --ino's are
kinematically accessible ($m_{\chi_2^0}\simeq m_{\chi^-_1}\simeq M_2$,
$m_{\chi_1^0}\simeq M_1$). The decays into $\chi_2^0$ and $\chi_1^+$ are
closed when $M_1\gsimeq 75$ GeV.

As discussed above, the  decay of $\sti$ into $\tau_{R(L)}$ has 100\%
branching ratio for $\sti=\st_{R(L)}$. $\brchi$ is however quite small if
$\sti \simeq \st_L$ as long as the other decay modes are kinematically
accessible, less than 20\% for $M_2\ll \msti$. This can be explained in the
limit $\msti\gg M_2$ as the ratios of couplings:
\beq
\gami &: \gamii : \gamci
\nonumber\\
for\  \st_L\ \ \   & = (a_{L1}^L)^2 : (a_{L2}^L)^2 : (b_{L1})^2
                  = \left(\frac{\tw}{\sqrt{2}}\right)^2 :
\frac{1}{2}: 1,
\nonumber\\
for\  \st_R \ \ \  &=(a_{R1}^L)^2:(a_{R2}^L)^2:0
                 = 1:0:0\label{e12}.
\eeq
{}From Eq.(\ref{e12}), one can see that $\sum_i Br(\chi_i^0 \tau)$ is less than
50\% and $Br(\chi_1^0)\sim$ $Br(\chi_2^0\tau)$$\tws$ for $\stl$. If $\sti$ is
$\st_R$, the modes into $\chi_2^0$ or $\chi^-_1$ have zero branching ratio, as
$\st_R$ is an SU(2) singlet.

In fig. 3, we also show the branching ratios into $\chi_1^0 \tau_R$ and
$\chi_1^0 \tau_L$ for $\sints=0.7$ (dotted lines). For $M_1<75$ GeV, the
branching ratios into $\chi_2^0$ or $\chi_1^-$ rise rather quickly reducing
the ratio into $\chi_1^0$ at the same time. This is again because the $SU(2)$
gauge coupling is larger than the $U(1)$ coupling. On the other hand,
the branching ratio into $\chi_1^0\tau_R$ is still more than 80\% of the total
$\chi_1^0\tau$ decay. Indeed, $\ptau ( \sti\rightarrow\chio\tau) $ for
$\chi_1^0 \simeq \tilde{B}$ is
\be
\ptau= \frac{(\sint a_{R1}^R)^2 - (\cost a_{L1}^L)^2}
{(\sint a_{R1}^R)^2 + (\cost a_{L1}^L)^2}
=\frac{4\sints -\costs}{4\sints+\costs}.
\label{e23}
\ee
The factor of 4 comes from the fact that $\tau_R$ has twice as large a
hypercharge as $\tau_L$. Due to this factor, if $\sint$ is close to one,
$\ptau$ has very mild dependence on $\thetat$, while $\ptau$ is quite
sensitive to the $\st_R$ component when $\sti$ is close to $\st_L$.

So far we have been discussing the case where  $M_1<\msti\ll\vert\mu\vert$.
How about the case $\vert\mu\vert<\msti$? In fig. 4a and fig. 4b, we show the
branching ratios of $\st_R$ and $\st_L$ for $\mu=-100$ GeV, $\msti=150$ GeV
and $\tanb=10$, varying $M_1$ from 50 GeV to 500 GeV. For $M_1\gg 100$ GeV,
$\chi_1^0$, $\chi_2^0$ and $\chi_1^-$ are higgsino-like, and  the decays  into
them are kinematically accessible, as $m_{\chi_1^0}$, $m_{\chi_2^0}$ and
$m_{\chi_1^-}\simeq |\mu|$. On the other hand, for $M_1<100$ GeV, $\chi_1^0$
has a large $\tilde{B}$ component. However, due to the relatively small value
of $|\mu|$, all $\chi_i$'s have non-negligible mixing between gauginos and
higgsinos, if $M_1$ is small.

In fig. 4a we show branching ratios of $\st_R$ decays. Solid lines show
branching ratios into the lightest neutralino. The solid line which decreases
as $M_1$ increases is for $Br(\chi_1^0 \tau_R)$, while the one which increases
as $M_1$ increases is for $Br(\chi_1^0\tau_L)$.  Dashed lines are for the sum
of the branching ratios into neutralinos, and the dot-dashed line is for
$Br(\chi_1^-\nu_{\tau})$. Branching ratios into the other final states are
negligible, and we did not show them on the plot.

For $M_1<100$ GeV, the bino component of $\chi_1^0$ decreases as  $M_1$
increases, while increasing the bino component of heavier states. Thus
$Br(\chi_1^0\tau_R)$ decreases, keeping $\sum_i Br(\chi_i^0\tau_R)$ about
constant. Increasing $M_1$ even more results in the kinematical suppression of
the decay into $\chi_3^0\simeq\tilde{B}$, reducing $\sum_i Br(\chi_i^0\tau_R)$
while increasing $Br(\chi_1^0\tau_R)$. However, for $M_1> 150$ GeV the
branching ratios into $\tau_R$ decrease monotonically, as the bino components
of $\chi_1^0$ and $\chi_2^0$ become smaller. Finally, in the limit where
$M_1\gg\vert \mu\vert$, only decays into $\chi_i^0\tau_L$ and
$\chi_i^-\nu_{\tau}$ survive. $Br(\chi_1^0\tau_L)\simeq$
$Br(\chi_2^0\tau_L)\simeq$ $1/2 Br(\chi_1^-\nu_{\tau})$ and
$Br(\chi_1^0\tau_L)\simeq 25$\% in this limit.

The branching ratios for $\st_L$ decay for the same value of $\mu$ and
$\tanb$, which are shown in fig. 4b, are quite different from those of $\st_R$.
For $M_1< \vert \mu\vert$, $Br(\chi_1^0\tau_L)$ (solid line) is very small,
while $Br(\chi_i^0 \tau_L)$ (dashed line) is larger than that of $Br(\chi_i^-
\nu_{\tau})$ (dot--dashed line). The branching ratios are also quite different
from those for $\mu=-300$ GeV shown in fig.2. This is due to a cancellation
between $N_{11}$ and $N_{12}$ in $a^L_{L1}$ (see Eq.(\ref{e6})), which
determines $Br(\chi_1^0\tau)$; the wino component $N_{12}$ is smaller than the
bino component $N_{11}$ and has the opposite sign, while the gauge coupling is
larger for the wino component, making the total $a^L_{1L} $  very small. For
$M_1\simeq 140$ GeV, these two components cancel exactly, leading to zero
branching ratio into $\chi_1^0\tau_L$. On the other hand, $N_{21}$ and
$N_{22}$ have the  same sign in this region, giving a large $a^{L}_{L2}$
coupling, and hence a large $Br(\chi_2^0\tau_L)$.

$Br(\chi_i^0\tau_L)$ reduces very quickly once $\vert M_1\vert> 100$ GeV. This
is because the decay into $\chi_3^0$ becomes kinematically suppressed, and
furthermore not only $a_{L1}^L$, but also $a_{L2}^L$ tends to be suppressed
due to a cancellation between the wino component and the bino component. The
cancellation also can be seen from Eq.(\ref{e13}), where the wino like and the
bino like components of the eigenvectors have opposite sign if $M_1, \
M_2\gg\vert \mu \vert$. As a result, the branching ratios into $\tau_R$ rise
rather quickly. Notice also that $Br(\chi_i^-\nu)$ is very small in this
region, as $\st_L$ does not couple to charged Higgsino. Therefore
$Br(\chi_1^0\tau_R) \simeq$ $Br(\chi_2^0\tau_R)\simeq$ 50\% in the limit
$M_1\gg \vert\mu\vert$.

There are several generic features which have been illustrated in fig. 3 and
4. First, if $M_1\ll \mu$, $\chio$ is bino--like and the branching ratio to
$\chio$ is dominant for $\str$, while it is rather small for $\stl$ as long as
wino--like $\chi^0_2$ and $\chi_1^-$ modes  are accessible (see fig. 3). On
the other hand, if the lightest neutralino is higgsino--like
($\vert\mu\vert\ll M_1$) and $\msti<M_1$ (which implies that decay modes into
gaugino--like --inos are kinematically forbidden), there is always a
substantial branching ratio into $\chio$, more than 25\% (50\%) for $\str$
($\stl$) (fig. 4). It should be noted, however, that the mass differences
between $\chio,\chi_2^0$ and $\chi_1^-$ are very small in this limit. For the
parameters shown in fig. 4, $\vert \mchi\vert=94.1$ GeV, $\vert
m_{\tilde{\chi_2^0}}\vert=106.6 $ GeV and $\vert m_{\chi_1^-}\vert=100.6$ GeV
for $M_1=400$ GeV. For such a small mass difference, the decays of $\chi_2^0$
and $\chi_1^-$ give very soft hadron jets or leptons, which may not be tagged.
Thus it is possible that we measure the sum of the decay modes
$\sti\rightarrow \chi_i^0\tau$ altogether and $\sti\rightarrow\chi_i^-\nu$ is
not detectable at all.  However, the relation between the parameters of  the
neutralino sector ($M_1,M_2,\mu,\tanb$)  and $\ptau$ is roughly unchanged,
even if we do the inclusive analysis. Finally, if $|\mu| \simeq M_1$, the
neutralinos and charginos are strong mixtures of gauginos and higgsinos.
$\str$ decays into $\chi_i^0\tau$ through their bino component, while $\stl$
decays into both $\chi_i^0\tau$ and $\chi_i^-\nu$ through their bino and wino
components. Interferences between gaugino components can be very important for
$\stl$. $\st_{R(L)}$ decays into $\tau_{R(L)}$, since the higgsino-(s)fermion
coupling is much smaller than the gaugino coupling unless $\tanb\gg 1$.

We should mention that production of $\tilde{\nu}_{\tau}$ may not be
distinguishable from $\sti$ production. $\tilde{\nu}_{\tau}$ production occurs
through $s$ channel $Z^0$ boson exchange. The production cross section in the
limit $s \gg m_Z$  can be obtained by replacing $1+\tws\rightarrow -1+\tws$
and taking $\sint=0$ and $\cost=1$ in Eq.(\ref{e4}); if $P_e=1$, the cross
section of $\tilde{\nu}_{\tau}$ production is equal to that for $\stl$. The
$\tilde{\nu}_{\tau}$ then decays into $\nu_{\tau}\chi_i^0$ or $\tau\chi_i^+$,
and the decay into $\chi_i^+$ gives $\tau$ + (2 jets or 1 lepton) + missing
momentum. Then the event signature is very close to that of $\sti \rightarrow
\tau \chi_i^0(i\ge 2)$ if the $\chi_i^+$ decays hadronically. Notice that
$\tilde{\nu}_{\tau}$ production always associates with $\sti$ production  if
the $\sti$ is close to $\stl$. This is because  the sneutrino soft breaking
mass parameter is common to that of $\stl$ due to $SU(2)\times U(1)$ symmetry;
$m_{\stl}\simeq m_{\tilde{\nu}_{\tau}}$ up to small D term effects. However,
the polarization of the $\tau$ lepton from $\tilde{\nu}_{\tau}$ decay is
left--handed (right--handed) for gaugino--(higgsino--)like $\chi_1^-$, thus
again the relation between $\ptau$ and $M_1,M_2,\mu,\tanb$ is
retained.\footnote{As will be discussed in sec. 4, $\ptau$ can be measured
only through the energy distribution of the decay products of the $\tau$
lepton. In order to be able to claim that $\tilde{\nu}\rightarrow\tau\chi_i^+$
decay does not disturb $\ptau$ information from $\tilde{\tau}\rightarrow
\tau\chi_j^0$ decay, $m_{\chi_i^+}\simeq m_{\chi_i^0}$ should be satisfied.
This is actually the case if $M_1\gg\vert\mu\vert$  or $\vert\mu\vert\gg M_2$,
as a wino--like neutralino and a wino--like chargino and also higgsino--like
neutralinos and a higgsino--like chargino are very close in mass.}

\section*{4) Measurement of $\ptau(\sti\rightarrow \tau\chi_1^0)$ }
In  this section, we discuss how to measure $\ptau(\sti\rightarrow\tau\chio)$.
We have already mentioned in the beginning of section 3 that the measurement
can be done through the energy distribution of the decay products of the
$\tau$ lepton. However, difficulties arise as the $\tau$ lepton does not have a
fixed energy; thus the energy distribution of the decay products depends not
only $\ptau$ but also on $\msti$ and $\mchi$.

The energy distribution of the $\tau$ lepton from the decay  $\sti\rightarrow
\tau\chi_1^0$  is flat between $\etaumax$ and $\etaumin$, where
\be
E_{\tau}^{\rm max(min)}
=\frac{\etauc\pm\ptauc\vstau}{\sqrt{1-\vstau^2}}.
\label{e16}\ee
Here $\etauc$ and $\ptauc$ are the  energy and the momentum of the $\tau$
lepton in the rest frame of $\sti$, and $\vstau$ is the velocity of $\sti$ in
the laboratory frame:
\beq
\etauc&=\frac{\msti^2+m_{\tau}^2-\mchi^2}{2\msti},
\nonumber\\
\ptauc&=\sqrt{\etaucs-m_{\tau}^2},
\nonumber\\
\vstau&=\sqrt{1-(2\msti/\sqrt{s})^2}.
\label{e17}\eeq
For $\se$ and $\smu$ production, the end points of the energy distribution of
the final state lepton can be used to determine $\mchi$ and $\msl$\cite{11}.
For the case of $\tilde{\tau}$, however, the $\tau$ lepton further decays into
$A\nu$, where $A= e\nu,\mu \nu, \pi,\rho,a_1...$. In the limit $E_\tau\gg
m_{\tau}$, $A$ keeps the original direction of the momentum of the $\tau$
lepton; however, the energy of $A$ is softer as some of the $\tau$ energy is
carried away by $\nu_{\tau}$.

As we have already mentioned, the energy distribution of the final state
particles strongly depends on $\ptau$. For illustrative purposes we discuss
the decay mode $\tau\rightarrow\pi\nu_{\tau}$ first. The energy distribution
of the $\pi^-$ arising from polarized $\tau$ leptons is very simple:
\be
\frac{1}{\Gamma_\tau}\frac{d\Gamma}{dz}
(\tau^{\pm}\rightarrow \pi^{\pm}\nu_{\tau})
=B_{\pi}\left[1\mp \ptau (2z-1)\right],
\label{e18}\ee
where $z\equiv E_\pi/E_{\tau}$ and $B_{\pi}\simeq 0.12$ is the branching ratio
of $\tau$ into $\pi$. One can see that the energy distribution is harder for
$\tau^-_R$ than for $\tau^-_L$. The  energy distribution of the $\pi$ coming
from the cascade decay of $\sti$ is then given by:
\beq\label{e19}
\frac{1}{\Gamma_{\sti}}\frac{d \Gamma}{dy}
&(\epem\rightarrow \sti\sti^*\rightarrow
\chio\tau^{-}\rightarrow \pi^{-}\nu_{\tau})
\nonumber\\
&={\cal F}\int^{x^{\rm max}}_{{\rm max}( x^{\rm min}, y )}
\frac{1}{x}\cdot \left[1+\ptau \left(\frac{2y}{x}-1 \right)\right]
\nonumber\\
=&\left\{\begin{array}{l l}
{\cal F}\left[(1- \ptau) \log\left(\frac{x^{\rm max}}{y}\right)
+ 2 \ptau\left( 1-\frac{y}{x^{\rm max}}\right)\right]
& {\rm for}\  y>x^{\rm min},
\\
\\
{\cal F}\left[\ (1- \ptau) \log\left(\frac{x^{\rm max}}{x^{\rm min}}\right)
+ 2 y \ptau \left(\frac{1}{ x^{\rm min}}-\frac{1}{x^{\rm max}} \right)
\right]&{\rm for}\  y<x^{\rm min},
\end{array}
\right.
\eeq
where $y=E_{\pi}/E_{\sti}=2E_{\pi}/\sqrt{s}$, $x^{\rm min}=
2\etaumin/\sqrt{s}$, $x^{\rm max}= 2\etaumax/\sqrt{s}$ and
\be
{\cal F}= \frac{1}{2 \vstau \ptauc}\sqrt{1-\vstau^2} B_{\pi}B_{\chio \tau}.
\label{e21}\ee

In fig.5 we show the energy distribution of the $\pi$ for $\ptau=\pm 1$ and
$0$, where $\mst=150$ GeV, $\mchi=100$ GeV and $\sqrt{s}=500$ GeV. For
$\ptau=+1$, the distribution is peaked at $y=x^{\rm min}\simeq 0.06$ and goes
to zero at $y=0$. On the other hand, the distribution is strongly peaked at
$y=0$ for  $\ptau=-1$. 48\% of the total events has  $y<x^{\rm min}$ for
$\ptau=-1$, while it is only 10\% for $\ptau=+1$.

Given the relatively hard $E_{\pi}$ spectrum peaked at $x^{\rm min}$, this
mode might be suitable to measure $x^{\rm min}$ and $x^{\rm max}$ if
$P_{\tau}\simeq 1$. If $x^{\rm min}$ and $x^{\rm max}$ can be measured, it is
easy to determine $\ptau$ and $\sigma(\epem\rightarrow\sti\sti^*)\cdot
Br(\sti\rightarrow\tau\chio)$. For example, one may count the number of events
with $y>x^{\rm min}$ and, say, $y>(x^{\rm min}+x^{\rm max})/2$, and carry out
a two parameter fit of $\ptau$ and $\sigma\cdot Br$. For the parameters used
in fig. 5, 52\% (90\%) of the events are in $y>x^{\rm min}$ for $\ptau=-1 \
(+1)$, while 4\% (22\%) of the events go above $(x^{\rm min}+x^{\rm max})/2$.
For ${\cal L}= 100fb^{-1}$, $\sigma(\epem\rightarrow \sti\sti^*)= 0.1$ pb and
$Br(\sti\rightarrow \chio\tau)=1$, more than one thousand $\pi^-$ events are
available for this analysis before any selection cuts have been applied.

For the above analysis, we needed to know $x^{\rm max}$ to determine `high'
and `low' energy samples. The precise measurement of this quantity turns out
to be  difficult for $\ptau \simeq -1$ due to the very small probability of
$\pi$ to get an energy near the kinematical maximum; for $\msti=150$ GeV and
$\msti=100$ GeV, only 1\% of the events have energy above $y>(x^{\rm min} +
2x^{\rm max})/3$ ($E_{\pi}>87$ GeV). Furthermore, if $x^{\rm min}$ is small,
its measurement will also be difficult. This is due to the huge background
from $\gamma\gamma\rightarrow \tau\bar{\tau}, q\bar{q}$ in the small $y$
region. The total cross section for the process $\gamma \gamma \rightarrow
\tau \bar{\tau}$ is about 7 pb.\footnote{There would also be a background from
beamstrahlung, which strongly depends on the machine design.} Imposing a
minimum visible energy cut of 50 GeV, in addition to other selection cuts, the
background can be reduced to less than 1 fb\cite{6}. The visible  energy cut
might be weakened, but still it would be very difficult to measure the shape
of the energy distribution over the $\gamma\gamma$ background in the small $y$
region. Notice that the detection efficiency of the $\pi$ depends strongly on
$P_{\tau}$ and $E_{vis}$. Together with the uncertainty of $x^{\rm max}$, this
makes the determination of $\ptau$ and $\sigma( \epem \rightarrow \sti \sti^*)
\cdot Br(\chi_0 \tau)$ quite difficult if one uses this $\tau$ decay mode
only. The problem of measuring $x^{\rm max}$ arises also in the decay mode
$\tau\rightarrow\mu (e)$, as $\mu (e)$ from $\tau$ decay has a very soft
energy spectrum both for $\ptau=\pm 1$.

Decay modes into heavier mesons,  which give multiple pions in the final
states, are more suitable to measure $x^{\rm max(min)}$. The $\tau \rightarrow
\rho\nu$ mode has a branching ratio of about 23\%, and the $\tau \rightarrow
a_1\nu$ mode has a branching ratio of about 15\%. $\tau_{L(R)}$ decays
dominantly into the longitudinal (transverse) component of $\rho$ or $a_1$,
and they tend to get most of the $\tau$ energy \cite{10}. The $\rho$ ($a_1$)
further decays into $2\pi \ (3\pi)$ states. Thus even when $\ptau=-1$, one
might be able to measure $x^{\rm max(min)}$ by studying the shape of energy
distributions of these multiple pion systems.

Even if $x^{\rm min}$ is not measurable due to backgrounds from $\gamma\gamma$
production, one might still be able to determine $\mchi$  from the production
of other sparticles. As we mentioned earlier, we can measure $\mchi$ through
the measurement of the end points of the energy distribution of $e(\mu)$ from
$\se$ ($\smu$) (if it is produced at the same beam energy). Another
possibility is to perform a threshold scan to determine $\msti$. Combining the
information of one of the two masses with $x^{\rm max}$, one can extract the
other mass, and thus $x^{\rm min}$. Once $x^{\rm min,max}$ are known, $\ptau$
can be extracted from the energy distribution of $\pi$ as described earlier.
$\ptau$ also can be measured independently by studying the distributions of
the difference of the energy between decay pions from $\rho$ and $a_1$.
Namely, transversely polarized $\rho$ favors equal splitting of the $\rho$
energy between the two decay pions, whereas longitudinally polarized $\rho$
leads to a large difference of the $\pi^-$ and $\pi^0$ energies. For $a_{1T}$,
all three pions have a tendency to share equally the energy of $a_1$. On the
other hand, $a_{1L}$ again favors configurations in which one or two of the
the pions are soft. Notice that, unlike the case of the decay mode
$\tau\rightarrow$$\pi\nu$, the dependence of the detection efficiency on
$\ptau$  and $ E_{vis}$ is not severe for these modes, as $\ptau$ mainly
affects the polarization of the vector mesons, rather than its energy
distribution. A detailed discussion of the energy distribution may be found in
\cite{10}.

There are several sources of background other than $W^+W^-$ boson production
which has been discussed earlier. The cross section  for $Z^0Z^0$ production
is about 0.15 pb for $\vert \cos\theta\vert<0.8$ and a  non-polarized electron
beam \cite{12}. Requiring one $Z^0$ boson to decay into neutrinos and the
other into a tau pair reduces this background to about 1.8 fb. The polarized
electron beam and a cut on the angle between the decay products of the two tau
leptons would make the background even smaller. $\sigma(\epem\rightarrow Z^0
h)$ is about 60 fb (for $m_h\sim 120$ GeV)\cite{12}. Demanding that the $Z$
boson decays into neutrinos and the Higgs boson goes into a tau pair reduces
this to about 1.8 fb. These backgrounds, including those coming from $W$ boson
production, can be calculated and subtracted reliablly from the signal events.
Notice that the polarization of $\tau$ leptons is zero for those coming from
Higgs bosons, and $-1$ for those coming from $W$ bosons. Those polarizations
affect the energy distribution of its  decay products, and have to be taken
into account for the estimation of the energy distribution of the background.
We finally remark that initial and final state radiation change the energy
distribution of the signal, which has not been included in this paper. MC
simulations both of signals and backgrounds are in progress\cite{17}.

\section*{5. Discussion and Summary}
What do we learn from the study of production and decay of $\sti$? If $\sti$
is the second lightest SUSY particle,  the branching ratio into
$\sti\rightarrow\chio\tau$ is 100\%. In this case, we can measure $\msti$,
$\mchi$, $\sigma(\epem\rightarrow \tau\bar{\tau})$ and $\ptau$ from the study
of $\sti$ production and decay. From the production cross section and
$m_{\tilde{\tau}_1}$, one can extract $\thetat$ as discussed in section 2. The
determination of $\thetat$ is very important as $A_{\tau}$ appears only in the
left--right mixing of $\sti$ or the $\stl\str H_1$ coupling. Once $\thetat$ is
known, the measurement of $\ptau$ would give one constraint on neutralino
mixing. Together with the measured value of $\mchi$, we get two constraints on
the four parameters in the neutralino and the chargino sector ($M_1, M_2, \mu,
\tan\beta$).

On the other hand, if the $\sti$ is not the second lightest SUSY particle, we
have to measure the branching ratios to other neutralinos and charginos to
determine the total $\sti$ pair cross section. As we have discussed in section
3, the $Br(\chi_1^0\tau)$ might be small if $ \sti\simeq \stl$ and decay modes
into heavier --inos are kinematically accessible. The ratio $Br(\chio\tau)$ /
$Br(\chi_i^0\tau)$ can be studied  by taking the ratio  $N(\tau\bar{\tau}$ +
missing  momentum) / $N(\tau\bar{\tau}$ + missing  momentum + jets or
leptons). The branching ratio into $\sti\rightarrow$ $\chi^-_j\nu_{\tau}$
would be more difficult to measure precisely, as one has to take the ratio
$N(\tau$ + missing momentum + jets or a lepton) / $N(\tau\bar{\tau}$ + missing
momentum + jets or  leptons). Because only one $\tau$ lepton is required for
the process where one $\sti$ decay into a chargino, the background from
$W^+W^-$ production is bigger, about 5 fb even with $P_e=1$. On the other
hand, in such a scenario pairs of charginos and heavier neutralinos can be
produced directly at the same collider. The study of these events, together
with the measured decay branching ratios of $\sti$, should give us much
information on the neutralino and chargino sector.

For the case with several SUSY particles which are close in mass to $\sti$ or
$\chio$, the analysis would be considerably more difficult. Even in such a
case, $\ptau$ gives a constraint to the combination of $\thetat$ and the
mixing matrix of the neutralinos ($N_{ij}$).

Several comments are in order. $b\ (\tilde b)$ and $t\ (\tilde t)$ also have
non-negligible Yukawa couplings to higgsinos, thus the polarization of bottom
and top quarks from $\tilde{b}$ and  $\tilde{t}$ decays depends on both squark
and neutralino mixing. However, the bottom quark quickly forms a hadronic
bound state, losing the information of the original polarization. A top quark
decays before making  a bound state. However, due to the large top mass
suggested by CDF and D0\cite{21}, at $\sqrt{s} = 500$ GeV the produced top
quark is too slow to allow a reliable measurement of its polarization. Thus
$\sti$ decay is the only place where one can possibly measure the polarization
of the final state fermion.

In this paper we discussed the production of $\sti$ at the proposed linear
colliders with $\sqrt{s}=500$ GeV. The same analysis in principle can be done
for $\epem$ colliders with smaller beam energy. At LEPII, however, $W^+W^-$
pairs are produced close to threshold, which makes precision studies of the
production of SUSY particles rather difficult. Besides, the integrated
luminosity is expected to be of order a few hundred pb$^{-1}$. Given that the
production cross section of $\sti\sti^*$ is about 0.3 pb for $\mstr=60$ GeV
and $\sqrt{s}=170$ GeV, a measurement the polarization of $\tau$ leptons
produced in $\sti$ decays seems to be very challenging.

Finally, we would like to discuss the mass spectrum of superparticles
($\sti,\chi_i^0,\chi_i^-$) in the minimal supergravity model, as this model
has been investigated in many articles recently. In this model, the mass
spectrum at the weak scale is expressed only by the 4 parameters $m, M, A(,
B)$ at the GUT scale; $m$ is a universal scalar mass, and $M$ a universal
gaugino mass. The relations to the weak scale quantities $m_R^2, m_L^2, M_1$
and $M_2$ are given by RG running from the GUT scale. The common mass
parameters at the GUT scale leads to equal soft slepton masses $m_R^2, m_L^2$
for $\tilde{l}_R$ and $\tilde{l}_L$ of all generations if the lepton Yukawa
couplings are negligible; numerically one has \cite{16}
\ben\label{e22}\beq
m_R^2&=m^2 +0.87 M_1^2,
\label{e22a}\\
m_L^2&=m^2 + 3 M_1^2,\label{e22b}
\\M_1&=\frac{5}{3}\tan^2\theta_W M_2=0.42M.
\label{e22c}\eeq\een
In particular, the relation $\mstr<\mstl$ follows from Eqs.(\ref{e22}),
suggesting that $\sti\simeq\str$. The relation between $M_1$ and $\mstr$ is
more complicated. If $m$ and the $D-$term are negligible compared to $M_1$,
$\mstr$ could be lighter than $M_1$, as the coefficient of $M_1^2$ in
Eq.(\ref{e22a}) is smaller than 1. In such a case, the decay modes into
gaugino--like states are closed.

The masses $m_{R(L)}$ are common to all sleptons if the $\tau$ Yukawa
coupling is negligible. However, $\sti$ can be lighter than the other sleptons
due to two effects. If $\tanb$ is big, the Yukawa coupling of $\tau$ is not
negligible so that $m_{\tilde\tau_{L,R}}$ at the weak scale get extra negative
contributions from the Yukawa interaction in the RG running from the GUT
scale, the effect being larger for $m_{\tilde\tau_R}$. Moreover, there can be
substantial mixing between $\stl$ and $\str$. Notice that this mixing tends to
increase with increasing $\tanb$ due to the term $\mu\cdot\tanb\cdot m_{\tau}$
in the off--diagonal element of the mass matrix of Eq.(\ref{e1})\cite{16}.
Mixing reduces $\msti$ even further, so that $\msti < m_{\tilde\tau_R} <
m_{\tilde{e}_R} \simeq m_{\tilde{\mu}_R}$ in minimal supergravity models with
$\tanb \gg 1$. If in addition $\msti< m_{\chi^-_1}$, $\sti$ might be the only
charged sparticle that is detectable at a given collider.

It is often said that the minimal supergravity model with radiative symmetry
breaking predicts $\vert \mu\vert>M_1$ for heavy top. This is due to the fact
that the Higgs mass parameters at the weak scale should not be too negative,
since otherwise the Higgs potential would be unbounded from below. One of
these mass parameters gets a large negative radiative correction coming from
(s)top loops when going down from the GUT scale. As the squared Higgs mass at
the GUT scale is the sum of squares of the soft mass parameter $m$ and the
higgsino mass parameter $\mu$, the large top Yukawa correction requires large
$|\mu|$ \cite{16,19}. In particular, for $m_t>160$ GeV or so, the lightest
neutralino is always bino--like \cite{15}; thus $\msti$ should be larger than
the bino--like neutralino mass $\simeq M_1$, to ensure that the lightest SUSY
particle is neutral. One would then expect $\sti$ to dominantly decay into
gauginos. However, this constraint would be drastically weakened if one allows
unequal masses for Higgs bosons already at the GUT scale. Recently such models
have been  constructed explicitly \cite{18}. Thus it is possible to find cases
where $\sti$ dominantly decays into higgsino--like neutralinos and charginos
even in a certain class of supergravity GUT models.

In summary, in this paper we proposed the measurement of the polarization of
$\tau$ lepton  that arises from the decay $\sti\rightarrow \chi_i^0\tau$. If
$\sti$ decays dominantly into $\chi^0_1\tau$, one can measure $\msti,\ \mchi$,
and $\thetat$, and get another constraint on the higgsino and gaugino mixing
of neutralinos by studying $\sti$ decay only. We would like to emphasize that,
if the production of other SUSY particles is also observed at future linear
colliders or hadron colliders, we could determine or even over-constrain all
SUSY parameters in the neutralino--chargino sector and the matter sector. This
would allow to simultaneously establish the existence of supersymmetry and
specify the GUT scale model that reproduces the observed mass spectrum at the
weak scale.

\section*{Acknowledgment}
I thank to K. Hagiwara, K. Fujii, T.Tsukamoto,
 Y. Okada, M. Drees and K. Hikasa
for helpful discussions, comments, and encouragements.  I would also
like to thank M. Drees for reading the manuscript carefully.  This
work is supported in part by the Grant-in-aid for
Scientific Research from the Ministry of Education, Science and
Culture of Japan (06740236).

\clearpage
\section*{Figure Captions}
\renewcommand{\labelenumi}{Fig.\arabic{enumi}}
\begin{enumerate}
\item 
The cross section for $\st_1\st^*_1$ production at an $\epem$ collider as a
function of the scalar tau mixing angle $\thetat$. We have taken $\msti=150 $
GeV and $\sqrt{s}=500$ GeV. For an unpolarized electron beam
$(P_{e^-}=0)$, the cross section depends very mildly on $\thetat$.

\vspace*{5mm}
\item 
Contours of constant polarization of the $\tau$ lepton arising from the decay
$\str\rightarrow\tau\chi_1^0$ in the plane spanned by $M_1$ and $\mu$. We have
assumed the $\tau$ lepton to be relativistic. The regions excluded by LEP
experiments are also shown in the figure. $\ptau$ decreases as $M_1$ increases
for fixed value of $\mu$.

\vspace*{5mm}
\item 
Branching ratios of the lighter scalar tau into a neutralino ($\chi^0_i$) plus
a $\tau$ or a chargino ($\chi^-_i$) plus a $\nu_{\tau}$, as a function of
$M_1$ for fixed $\msti=150$ GeV, $\mu=-300$ GeV, $\tan\beta=10$ and
$\sqrt{s}=500$ GeV; $\chio\simeq\tilde{B}$, $\chi_2^0\simeq\tilde{W^3}$ and
$\chi_1^-\simeq \tilde{W}$ for the parameters shown in this figure. If $\sti$
is $\str$, $\sti$ decays into $\chio \tau_R$ dominantly (the upper solid
line), while $\stl$ decays into $\chio\tau$, $\chi_2^0\tau$  and
$\chi_1^-\nu_{\tau}$ as long as these decay modes are kinematically
accessible.

\vspace*{5mm}
\item 
a) Branching ratios of $\str$ into $\chi_i^0\tau$ and $\chi_i^-\nu_{\tau}$ for
$\mstr=150$ GeV, $\mu=-100$ GeV and $\sqrt{s}=500$ GeV, varying $M_1$ from 50
GeV to 500 GeV. For $M_1\gg\vert \mu\vert$, $\chio$, $\chi_2^0$ and $\chi_1^-$
are dominantly higgsinos.

b) Branching ratios of $\stl$ into $\chi_i^0\tau$ and $\chi_i^-\nu_{\tau}$
for the same parameters as in Fig. 4a.

\vspace*{5mm}
\item 
Energy distribution of $\pi$ from the $\st$ cascade decay $\st
\rightarrow\chi_i^0\tau$ and $\tau\rightarrow\pi\nu_{\tau}$ at a
$\epem$ collider; $\sqrt{s}=500$ GeV, $m_{\tilde{\tau}}=150$ GeV
and $\mchi=100$ GeV. The three curves are for different $\tau$ polarization
$\ptau=1,0,-1$.
\end{enumerate}
\end{document}
\bye